# Cosmic ray radiography of a human phantom

Christopher Morris, John Perry, and F. E. Merrill

*Los Alamos National Laboratory, Los Alamos, NM, USA 87545.*

**Abstract.** Cosmic ray muons, that reach the earth's surface, provide a natural source of radiation that is used for radiography. In this paper, we show that radiography using cosmic radiation background provides a method that can be used to monitor bulk aspects of human anatomy. We describe a method that can be used to measure changes in patients as a function of time by radiographing them using cosmic-ray muons. This could provide hourly readouts of parameters such as lung density with sufficient sensitivity to detect time changes in inflammation of the lungs in e.g. Covid patients.

## Introduction

High energy hadronic interactions between high energy cosmic rays and the upper atmosphere produce showers of energetic pions that decay into muons. The leptonic muons have a relatively long mean free path, $l = \dfrac{1}{\beta \gamma \tau c}$, where $\beta = v/c$, c is the speed of light, $\tau$=2.2 μs is the muon life time, and $\gamma = \sqrt{\dfrac{1}{1-\beta^2}}$, and many are energetic enough to make it to the earth's surface.

In the recent COVID-19 pandemic there have been about 200 million deaths worldwide with many more expected. This disease results in many hospitalizations for times of up to several weeks. In severe cases COVID-19 infects the lungs, causing a buildup of pus and fluids[1] that is visible in CT scans and in radiographs references [2, 3].

We propose that the lungs of hospitalized patients can be monitored using cosmic ray muons. A schematic of such a monitoring system is shown in Figure 3. Muon trackers above and below the bed measure the trajectories of incident and outgoing muons as the pass through the patient. The angular distribution of the transmitted muons allows the measurements of the densities in the imaging.

We estimate that the impacted areas observed in the CT scans in figure 1 amount to more than 10% of the path length through the chest.

Using:

$$\frac{\Delta t}{t} \approx \frac{1}{\sqrt{N}} \qquad (0.1)$$

where $t$ is the measured thickness and $N$ is the incident fluence of cosmic ray muons in an area of interest. Exposure times on the order of an hour, assuming a flux of 1 muon/cm²/minute, can provide significant measurement of fluid buildup in the lungs with 1 cm resolution. Additional information can be obtained by measuring the stopping rate of muons.[4]

Here we examine the possibility of a similar analysis by combining mass, cosmic ray transmission, and cosmic ray multiple scattering. Transmission (or stopping) imaging with cosmic rays is somewhat different from point source x-ray imaging in that both the intensity and the direction of the cosmic rays can be measured. The trajectory information can be used to generate a focused transmission image at any distance from the detector.

Coulomb multiple scattering also contains material information. The angular distribution, neglecting logarithmic terms, is given by:

$$\frac{dN}{d\theta} = \frac{1}{2\pi\theta_0^2} e^{-\frac{\theta^2}{2\theta_0^2}} d\Omega$$

$$\theta_0 = \frac{14.1}{p\beta}\sqrt{\frac{L}{X_0}}$$

$$X_0 = \frac{K}{A}\left[Z^2\{L_{rad} - f(Z)\} + ZL'_{rad}\right]$$

Here dN/dθ the polar angular distribution of scatted particles is Gaussian with a width, θ₀,. A material can be characterized by its radiations length, $X_0$, where Z and A are its atomic length and Mass respectively, $K, L_{rad}, f(Z),$ and $L'_{rad}$ are parameters described in [5]

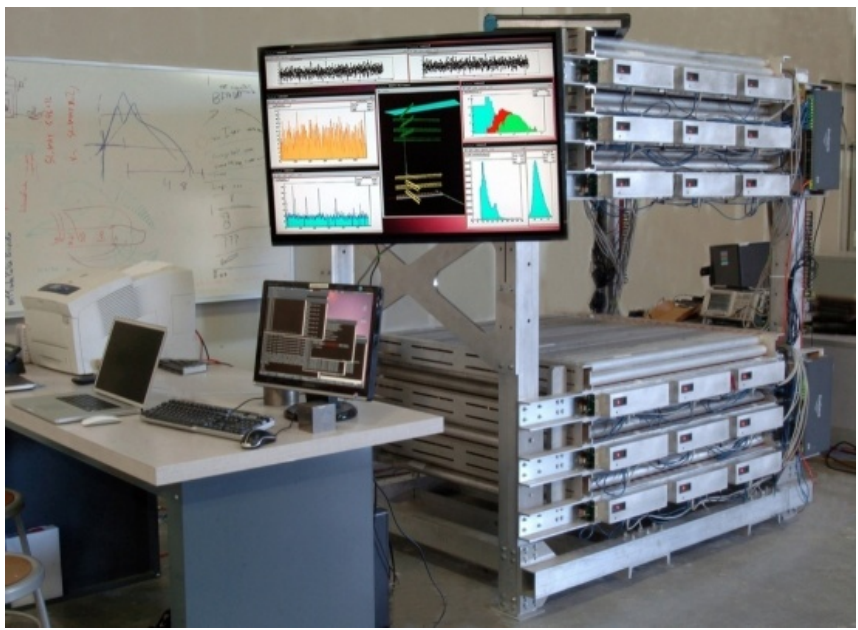

Figure 1) Photograph of the MMT. Objects for study were placed in the approximately two-feet (60 cm) gap between the two detector "supermodules".

## Cosmic ray imaging

The energy loss of muons ionizes and excites electronic states in material trough which they pass. This allows positions of muon tracks to be measured in an assortment of different kinds detectors. We have used drift tube detectors where the muon tracks are measured with ~0.5 mm precision in proportional tubes by using the timing information. Placing detectors above and below an object enables three different types of radiography, transmission, stopping, and multiple scattering. A photograph of the muon tracking detectors used here is shown as Figure 1A plot of the energy spectrum for overhead muons at sea level is given in Figure 3. Over a wide range of momentum, the energy loss for cosmic ray muons varies only logarithmically with momentum and is approximately proportional to the electron density, Z/A. Muon stopping can be understood as the shifting of the spectrum shown in Figure 2 to the left, with the loss of muons with energies below zero. The details of the MMT and the algorithms used here are given elsewhere. [4, 6]

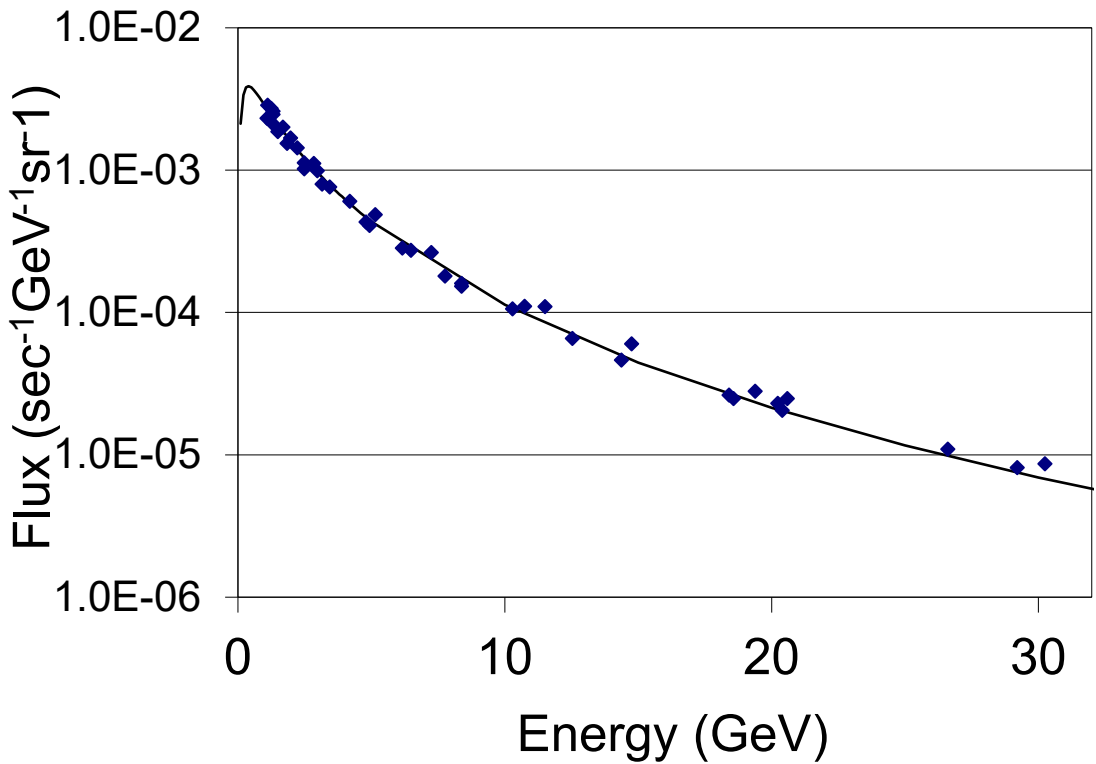

Figure 2) Spectrum of vertical cosmic ray flux at sea level. Solid symbols are the data.[9] The line is a parameterization.

## Transmission radiography

We have measured the transmission through the phantom. The transmission image was obtained by making a ratio of the image of transmitted cosmic rays with the target in place to an image with no target. The bottom and top tracks for transmitted trajectories had to intersect in a horizontal plane at the center of the object to within a radius of 1 cm. The resulting radiographs are shown in Figure 3.

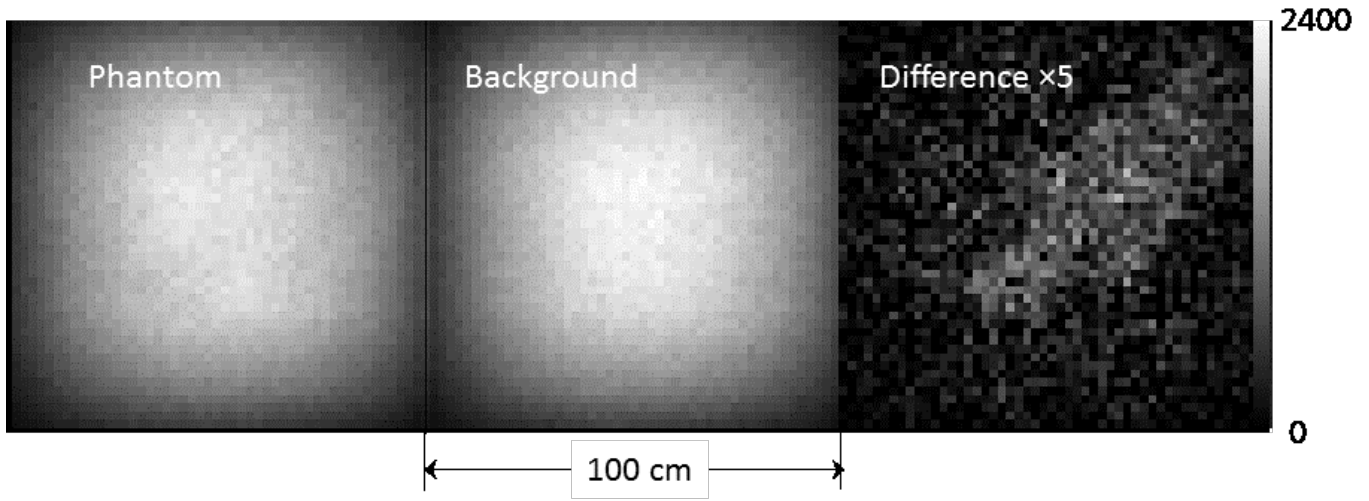

Figure 3) Transmission radiography.

## Stopped track radiography

Another type of radiography uses stopped tracks. An image made by projecting tracks that enter the top but are not detected in the bottom detectors is shown in Figure 4.

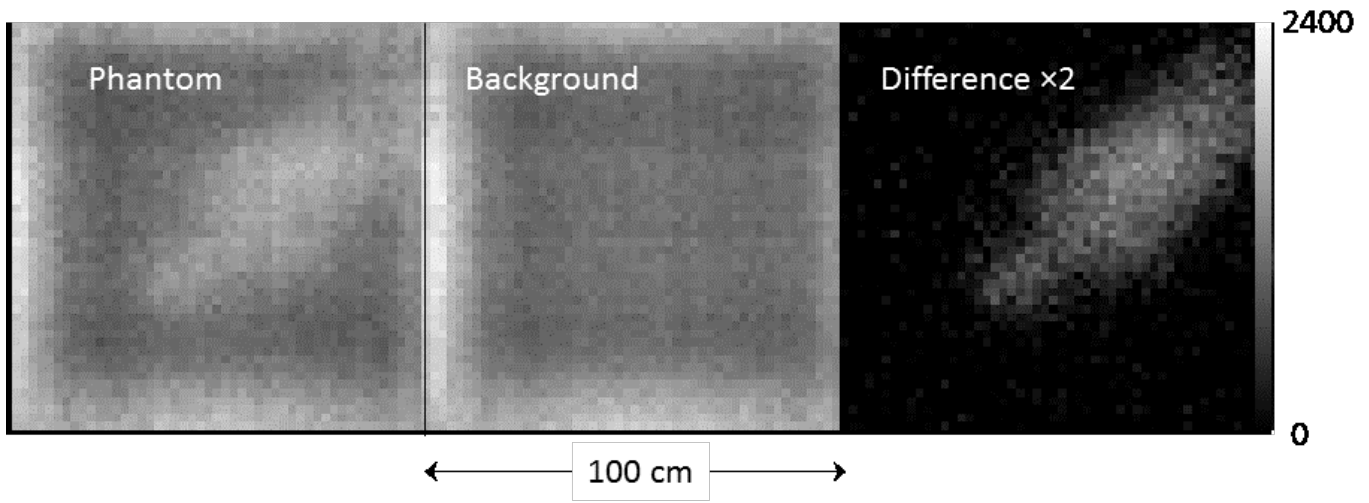

Figure 4) Left) Phantom Middle) background Right) Difference. The grey scale is linear in counts/4 cm$^2$ bin/24 hours.

## Multiple scattering radiography

Multiple scattering radiography is performed using the differences in angle (scattering) between incoming and outgoing muon trajectories to infer the thickness of an object measured in radiation lengths. In figure 3 we show an image of the human phantom obtained from multiple scattering radiography from the same data set used for the stoping and transmission images shown above. Although it is difficult to make a quantitative analysis. A clear difference between the two images is apparent. The skeletal structure shows more contrast relative to the body in the scattering image than in the stopped image. This is because of the extra power of Z in the computation of radiation length relative to stopping power enhances scattering in the higher-Z components of bone relative to tissue.

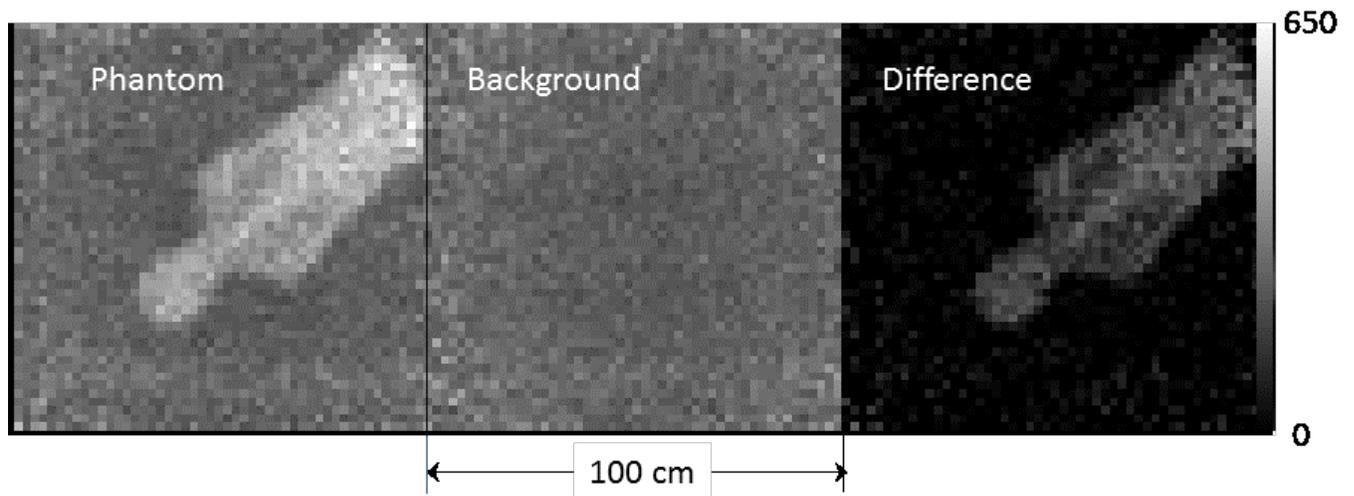

Figure 5) left) multiple scattering radiograph of the phantom, center) empty scene, right) the difference. The exposure was 24 hours.

## Conclusions

We have used experimental data collected with the Los Alamos Mini Muon Tracker to examine a human phantom. A 24-hour exposure was sufficient to obtain images using transmission, stopping and multiple scattering. In each case, the signal to background was ~1:1. The areal mass of detectors we used, dominated by the Aluminum drift tubes, was comparable to the phantom. Lower mass higher efficiency detectors, such as wire chambers, could provide higher quality images in considerably shorter exposure times with much better signal to noise. Changes in non-ambulatory patients may be able to be continuously monitored using only background radiation.

## Acknowledgements

This work was supported in part by the Los Alamos Laboratory Directed Research and Development (LDRD) office and part under the auspices of the U.S. Department of Energy under Contract DE-AC5206NA25396.